\documentclass[a4paper]{jpconf}

\usepackage{epsfig}
\usepackage{graphicx}

\def\J#1#2#3#4{#1 {\it #2} {\bf #3} #4}

\def\PRB{{\it Phys. Rev.} B}

\def\PRM{\it Rev. Mod. Phys.}

\def\be{\begin{equation}} \def\ee{\end{equation}}
\def\bea{\begin{eqnarray}} \def\eea{\end{eqnarray}}
\def\PRB{Phys. Rev. B}

\newcommand{\ket}[1]{| #1 \rangle}
\newcommand{\bra}[1]{\langle #1 |}

\begin{document}
\title{Coulomb correlations of a few body system of spatially
  separated charges}

\author{Ian Mondragon-Shem and Boris A. Rodriguez$^*$}
\address{Instituto de F\'{i}sica, Universidad de Antioquia, AA 1226,
  Medell\'{i}n, Colombia} \ead{$^*$banghelo@fisica.udea.edu.co}
\author{Francisco E. Lopez} \address{Centro de Investigaciones,
  Instituto Tecnol\'ogico Metropolitano, Calle 73 No 76A -354,
  Medell\'{i}n, Colombia}

\begin{abstract}
A Hartree-Fock and Hartree-Fock-Bogoliubov study of a few body system
of spatially separated charge carriers was carried out. Using these
variational states, we compute an approximation to the correlation
energy of a finite system of electron-hole pairs. This energy is shown
as a function of the Coulomb coupling and the interplane distance. We
discuss how the correlation energy can be used to theoretically
determine the formation of indirect excitons in semiconductors which
is relevant for collective phenomena such as Bose-Einstein
condensation (BEC).
\end{abstract}

\section{Introduction}
The physical properties displayed by charge carriers in spatially
confining settings could have a diverse range of potential
applications relevant to technological developments in the near
future. Systems with confined charge carriers such as semiconductor
quantum wells, quantum wires and quantum dots, will likely have an
important impact in the development of quantum computation and quantum
information \cite{Nielsen} and in producing new technological devices
in electronics, spintronics and optoelectronics \cite{Zutic}. In order
to take advantage of this important practical potential, it is
necessary to have a thorough theoretical understanding of the physics
of such systems. 

In particular, it is well known that the physics of electron-hole
pairs in quantum systems, such as quantum wells and quantum dots, is
fundamentally affected by Coulomb correlations (see, for example,
\cite{Alatalo} and \cite{Eisenstein}). As a consequence, a complete
theoretical study of the physical properties of charge carriers in
confining settings must take into account the very important
long-ranged Coulomb interaction.


With this in sight, we focus our study on a few-body system of
electron-hole pairs confined to vertically stacked parabolic quantum
dots. This type of system has been studied recently in the theory of
quantum dot molecules (e.g. \cite{Saikin}, \cite{Doty}). Our calculations will involve twelve
particles: six electrons and six holes, for a total of six
electron-hole pairs. It is clear that such a system is a long way from
the $O(10^2)$ pairs of real systems. However, the correlations that
arise here should give us a first reasonable insight into the real
physical system. 
\begin{figure}
\begin{minipage}{7cm}{\begin{center}\includegraphics[scale=0.27]{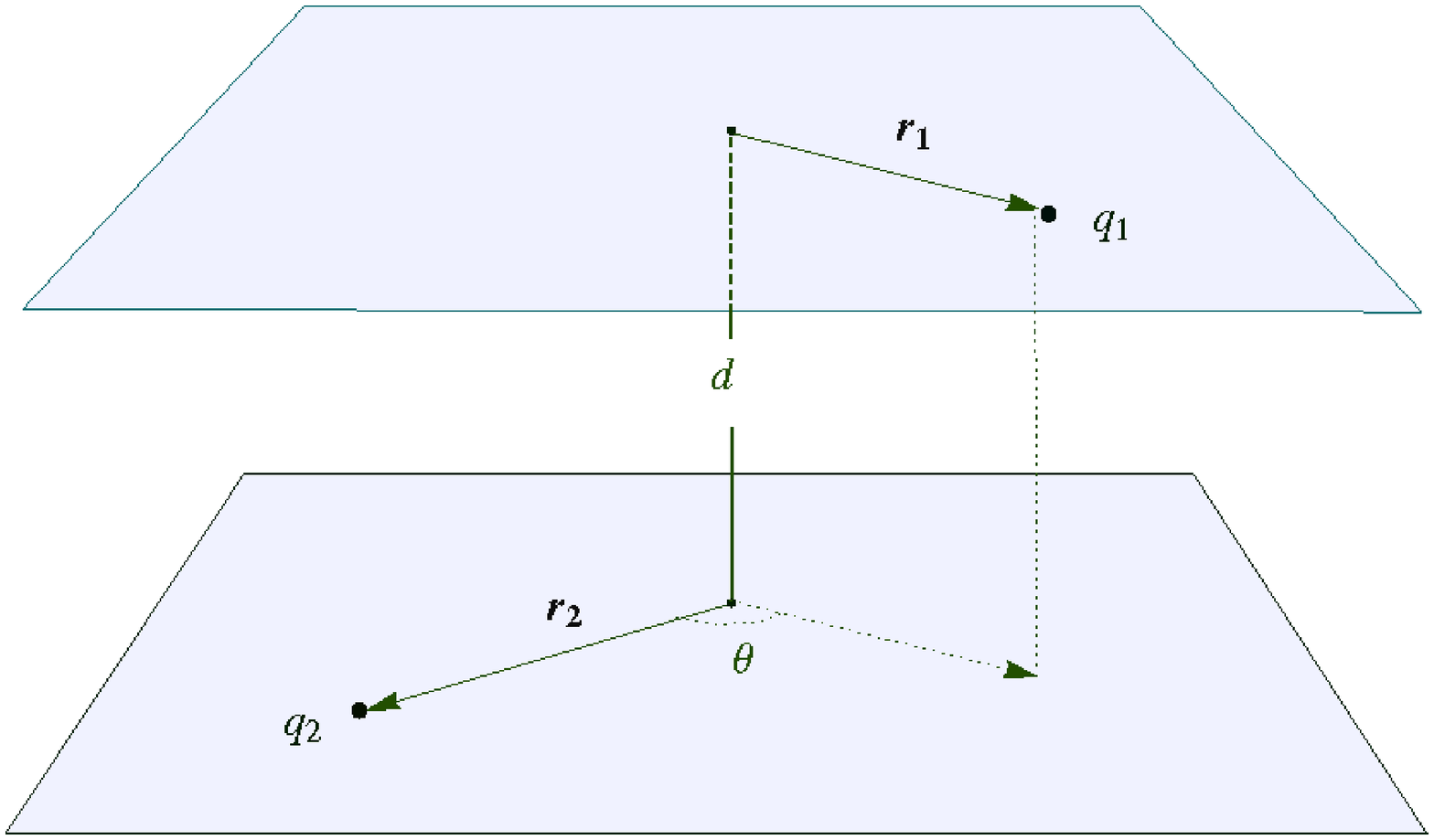}\end{center}}\end{minipage}
\begin{minipage}{8cm}{\textbf{Fig 1:} Schematic of our system with two charge carriers $q_1$ and $q_2$. They are in two different planes, spatially separated by a vertical distance $d$. The angle $\theta$ is the angle between the position vectors $\mathbf{r_1}$ and $\mathbf{r_2}$.}\end{minipage}
\end{figure}\label{system}

\section{Theoretical framework}

We compute the ground state energy of the system of electron-hole
pairs using two variational approximations used in finite system
studies: the Hartree-Fock (HF) state and a
Hartree-Fock-Bogoliubov-type (HFB) state. These states differ in that
the HF state fails to provide the so-called correlation energy of the
few-body system, whereas the HFB state is able to yield a first
approximation this energy. With this information, we will be able to
assess under which conditions the correlation energy due to
electron-hole pair formation is the greatest i.e. to assess when the
formation of spatially indirect excitons is favoured.

To achive our goal, it is necessary to have an efficient way of
computing the representation of the Coulomb interaction in the $2D$
harmonic oscillator basis. The Coulomb matrix elements read:
$\bra{ij}\hat{V}(d)\ket{kl}=\bra{ij}\frac{1}{\sqrt{\mathbf{r}_{12}^2+d^2}}\ket{kl}$,
where $d$ denotes the separation distance (see Fig.(\ref{system})) and
the indices in the bras and kets label the 2D harmonic oscillator
eigenstates . We have been able to compute the matrix elements
\cite{IFB} by expanding the integrals in terms of a single infinite
series followed by a series acceleration algorithm to achieve rapid
convergence.

The hamiltonian of the electron-hole pairs in this system is given by
\begin{small}
\begin{eqnarray}
H&=&\sum_i\{\epsilon^{(e)}_i
e^{\dagger}_ie_i+\epsilon^{(h)}_ih^{\dagger}_ih_i\}+\beta_{ee}
\sum_{ijkl}\bra{ij}V(0)\ket{kl}e^{\dagger}_ie^{\dagger}_{j}e_{l}e_k+\nonumber\\&&+\beta_{hh}
\sum_{ijkl}\bra{ij}V(0)\ket{kl}h^{\dagger}_ih^{\dagger}_{j}h_{l}h_k
-\beta_{eh}
\sum_{i\overline{j}k\overline{l}}\bra{ij}V(d)\ket{kl}e^{\dagger}_ih^{\dagger}_{\overline{j}}h_{\overline{l}}e_k.
\end{eqnarray}
\end{small}
\noindent
The $\beta_{ee}$, $\beta_{hh}$ and $\beta_{eh}$ parameters are dimensionless quantities that measure the
relative strength of the Coulomb interaction and the harmonic
potential: $ \beta_{xx}=\left(\frac{m e^4}{\epsilon_{xx}^2\hbar^2}/(\hbar
\omega_o)\right)^{1/2}$.  In this work, we take
$\beta_{ee}=\beta_{hh}=\beta_{eh}=\beta$, because we only seek to
understand the immediate physical effect of having spatial separation
between the charge carriers.  A more realistic calculation will need
to take into account the fact that the $\beta$ factor dependends on
the dielectric constant of the materials involved in the semiconductor
and, thus, their values are going to be different in general.  Through
out this work, we take the energy scale to be the harmonic potential
energy $\hbar \omega$ and the length scale to be the harmonic
oscillator length $l=\sqrt{\frac{\hbar}{m \omega}}$.

\subsection{The Hartree-Fock ground state}
In the first step of our calculation, we minimized the average energy
using a slater determinant in order to compute the ground state energy
of the few body system. The HF state is given by: $
\ket{\phi_{HF}}=\prod_{\mu=1}e^{\dagger}_{\alpha_\mu}\prod_{\nu=1}h^{\dagger}_{\beta_\nu}\ket{0}$.
As is well known, the Hartree-Fock energy provides the best energy for
uncorrelated single particle wave functions. Deviations from this
result will yield information of the correlations beyond a single
particle picture of the system. This transition to a correlated state
will involve, in a first approach, two-particle correlations between
electrons and holes, thus generating a picture of bound electron-hole
pairs.

\subsection{The BCS ground state}

In order to include correlation effects, in the second step of our
calculations we made use of the BCS (Bardeen-Cooper-Schrieffer) state,
given by:
$\ket{\phi_{BCS}}=\prod_i(u_i+v_ie^{\dagger}_ih^{\dagger}_{\bar{i}})\ket{0}$
which is one of a various types of Hartree-Fock-Bogoliubov correlated
variational ground state approximations.  Such a state introduces pair
correlations between electrons and holes from the outset. The
coefficients $v_i$ are the variational parameters in the system and
are a measure of the likelihood for an $i^{th}$ correlated pair to
appear in the system.


\section{Results and discussion}

\begin{center}
\begin{figure}
\centering
\begin{minipage}{7cm}\centering \includegraphics[scale=0.55]{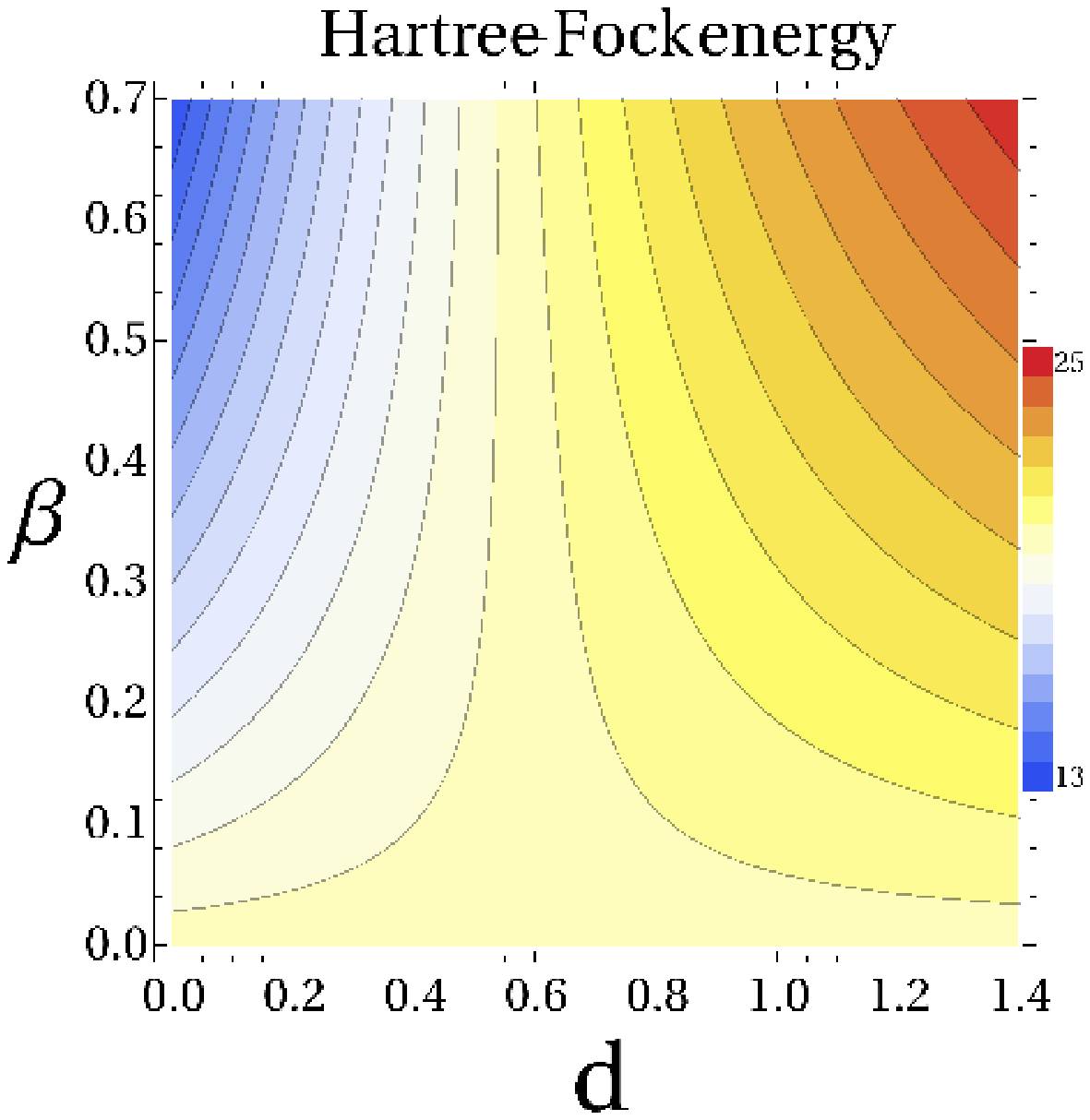}\end{minipage}
\begin{minipage}{7cm}\centering \includegraphics[scale=0.55]{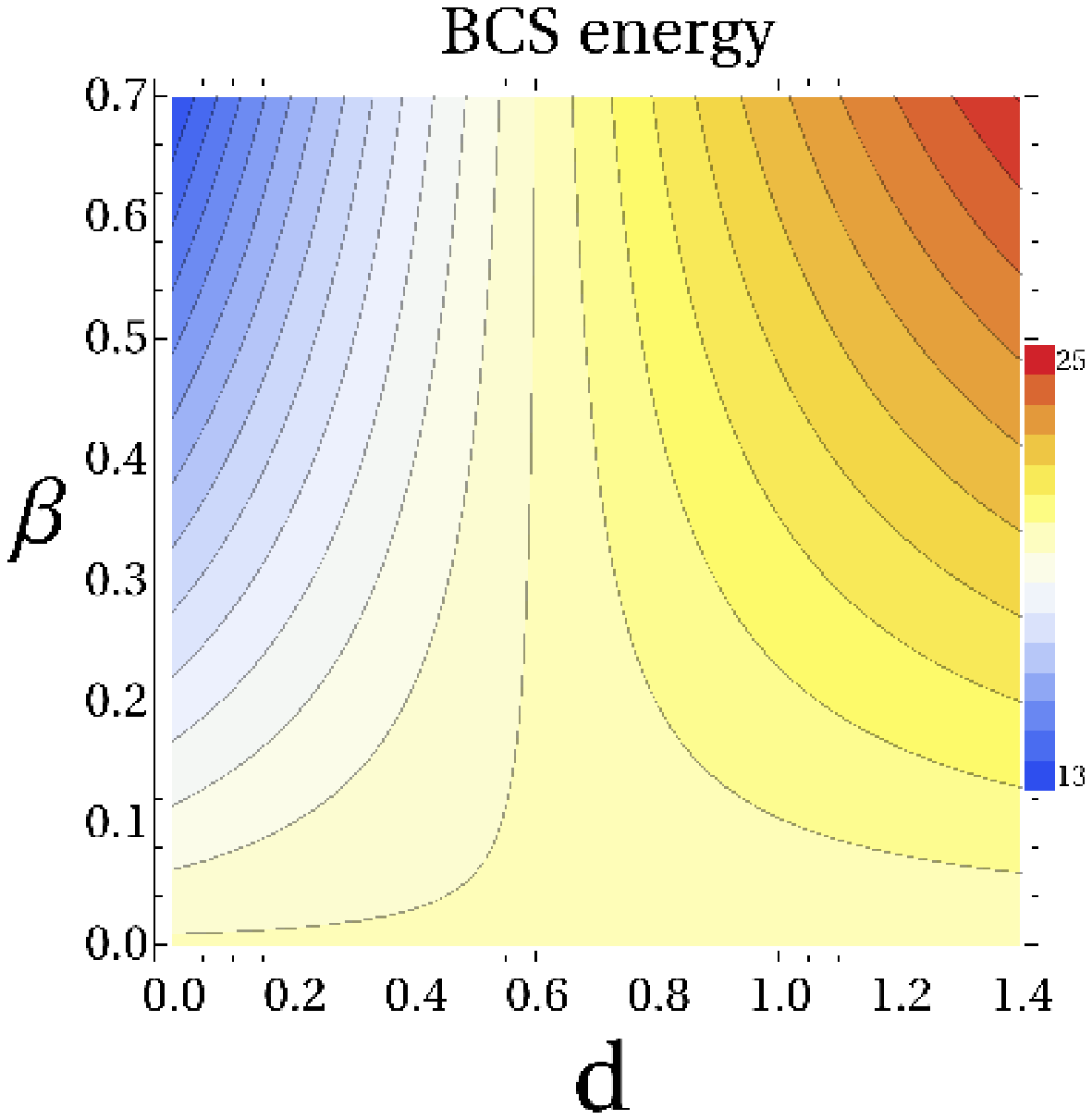}\end{minipage}
\caption{Contour plots of the HF and the BCS ground states in the
  $(\beta,d)$ parameter space.}\label{EBH}
\end{figure}
\end{center}

In Fig.(\ref{EBH}), we show the HF and BCS energies in the parameter
space $(\beta,d)$. Both of these plots tell us how, by a suitable
combination of the Coulomb coupling $\beta$ and the interplane
distance, we can go from energies below the energy of the
noninteracting particles to energies well above this value. The
appearance of the blue region shows that the attractive interaction
between electrons and holes is dominant over that of the repulsive
electron-electron and hole-hole ones, when $\beta$ increases or when
$d$ decreases. However, as we separate the planes, the electron-hole
energy gradually loses its predominance until we achieve a state in
which the contribution from repulsive and attractive interactions
cancels out, and we are left with the energy value of the
noninteracting system.

Note that the contour level corresponding to the particular situation
of equal contributions from attractive and repulsive energies is given
approximately by a small region of constant distances,
\textit{independent of the Coulomb coupling}. This region is shifted
to the right in the BCS plot, with respect to that in the Hartree-Fock
plot.

\begin{center}
\begin{figure}
\centering
\begin{minipage}{7cm}\centering\includegraphics[scale=0.22]{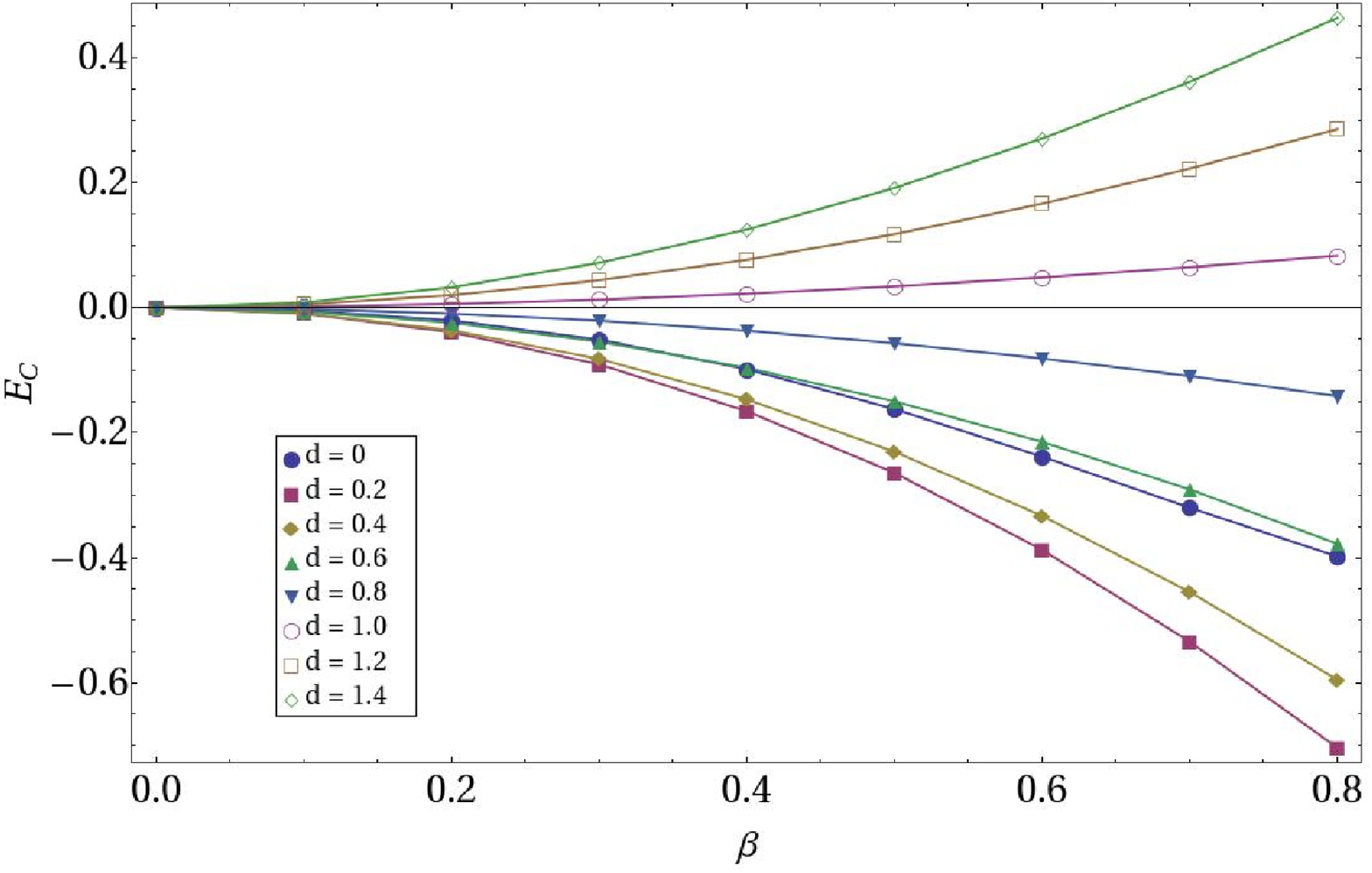}\end{minipage}
\hspace{0.7cm}
\begin{minipage}{8cm}\centering \includegraphics[scale=0.53]{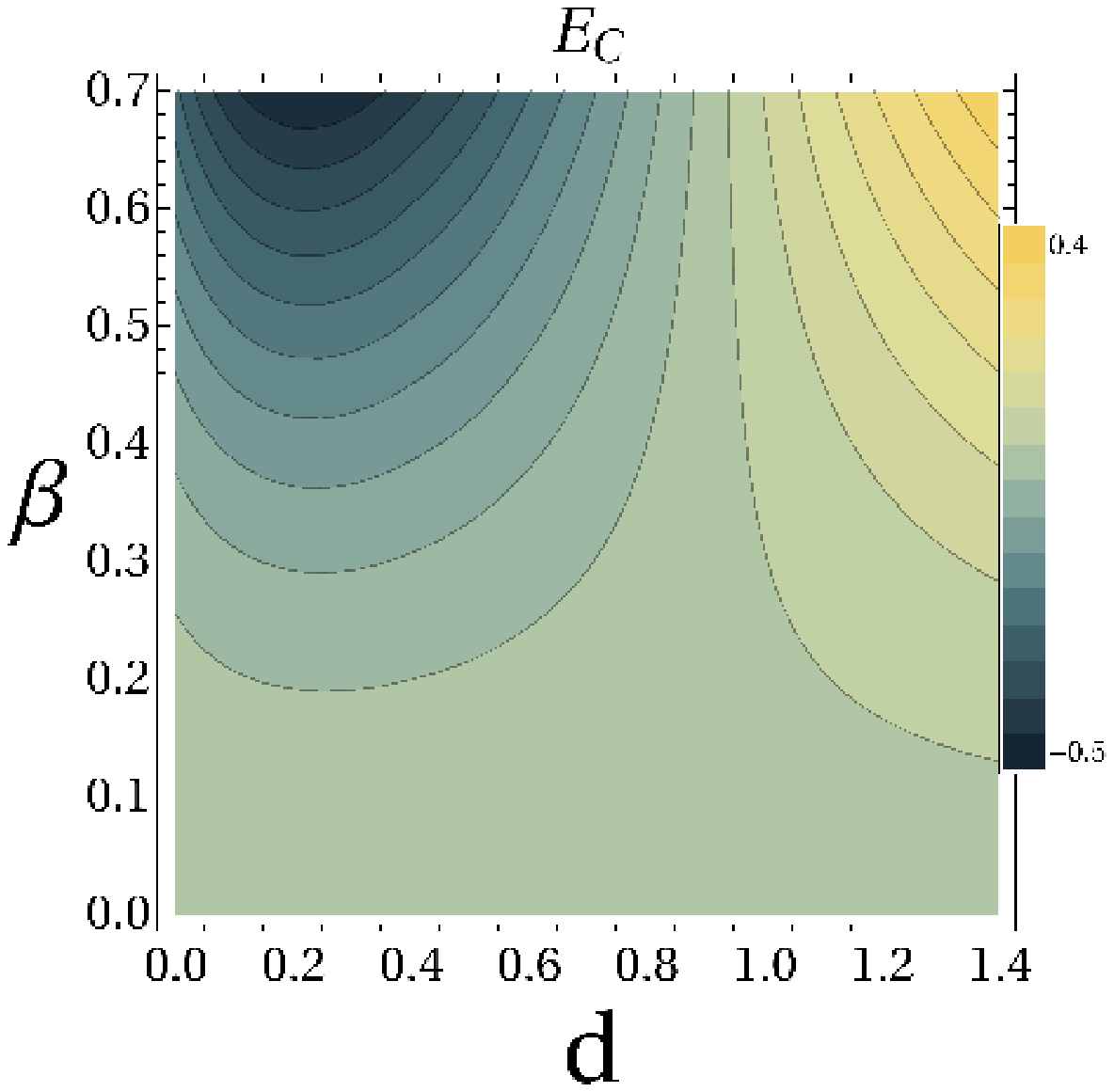}\end{minipage}
\caption{(\textbf{Left}) Correlation energy $E_{C}$ for various
  values of the distance $d$. (\textbf{Right}) Contour plot of the
  correlation energy in the $(d,\beta)$ plane.}\label{EC}
\end{figure}
\end{center}

The correlation energy, which is approximately given by $E_C(d,\beta)
\approx E_{BCS}(d,\beta)-E_{HF}(d,\beta)$, is shown in Fig.(\ref{EC});
both plots in this figure provide different views of the correlation
energy. We arrive at two important conclusions from the behaviour of
$E_C(d,\beta)$.
First, the system electron-hole pairs show the strongest excitonic
behaviour at a non-trivial value of interplane distance. This can be
understood as follows: when $d=0$, the interactions between the
particles are effectively screened since they all exist in the same
space i.e. in the two-dimensional plane. However, when there is a
finite distance between the planes, the screening between electrons
and holes is reduced and, thus, the electron-hole attractive Coulomb
interaction is reinforced. This enhanced attraction effect competes
with the electron-electron and hole-hole repulsive interactions and
with the suppression of the Coulomb interaction due to the finite
distance between the electrons and holes. The attractive interactions
eventually lose this competition when the repulsive interaction
dominates the ground state of the system, which happens at a critical
distance, as can be noted in the plot of the correlation energy.

This last physical turning point brings us to the second important
feature of the correlation energy: at the aforementioned critical
distance, the energy of the BCS state looses its superiority over the
Hartree-Fock energy, since the BCS state energy becomes larger than
that of the Hartree-Fock state. This means that the system is no
longer suitably described through correlated pairs i.e. the system has
suffered a structural change of quantum state, which is reminiscent of
the quantum phase transitions that occur in large many-body
systems. This transition point, together with the minimum of the
correlation energy for finite $d$, are important pieces of information
for understanding the BEC state of indirect excitons in
semiconductors.

\ack The authors acknowledge partial financial support from the CODI-
Universidad de Antioquia and the Centro de Investigaciones- Instituto
Tecnol\'{o}gico Metropolitano de Medell\'{i}n. I.M.S. is grateful for
useful discussions with J. L. Sanz on the Hartree-Fock method.


\section*{References}
\begin{thebibliography}{9}
\bibitem{Nielsen}M. A. Nielsen and I. L. Chuang. Quantum Computation
  and Quantum Information, Cambridge University Press 2000.
\bibitem{Zutic}I. Zutic, J. Fabian and S. Das Sarma,
  \J{2004}{\PRM}{76}{323}.
\bibitem{Alatalo} M. Alatalo, M.A. Salmi, P. Pietilainen and Tapash
  Chakraborty, \J{1995}{Phys. Rev. B}{52}{7845}.
\bibitem{Eisenstein} J. P. Eisenstein, G.S. Boebinger, L.N. Pfeiffer,
  K.W. West and Song He, \J{1992}{Phys. Rev. Lett.}{68}{1383}. 
\bibitem{Saikin} S. K. Saikin, C. Emary, D. G. Steel and L. J. Sham,
  \J{2008}{\PRB}{78}{235314}.
\bibitem{Doty} M. F. Doty, M. Scheibner, A. S. Bracker,
  I. V. Ponomarev, T. L. Reinecke, and D. Gammon,
  \J{2008}{\PRB}{78}{115316}.
\bibitem{IFB} I. Mondragon-Shem, F. E. Lopez and B.A. Rodriguez. Submitted.
\end{thebibliography}
\end{document}